\documentclass[runningheads]{llncs}
\usepackage[T1]{fontenc}
\usepackage{array}
\usepackage{cite}
\usepackage{amsmath,amssymb,amsfonts}
\usepackage{algorithmic}
\usepackage{subfigure, float}
\usepackage{graphicx}
\usepackage{textcomp}
\usepackage{xcolor}
\usepackage{url}
\usepackage{scrextend}
\usepackage[linesnumbered,ruled,vlined]{algorithm2e}
\begin{document}
\title{RSET: Remapping-based Sorting Method for Emotion Transfer Speech Synthesis}
%
%
\author{Haoxiang Shi\inst{1,2 \dagger}, Jianzong Wang\inst{1 \dagger}, Xulong Zhang\inst{1}\thanks{Corresponding author: Xulong Zhang (zhangxulong@ieee.org)\\$\dagger$ Equal Contribution}, Ning Cheng\inst{1}, Jun Yu\inst{2} and Jing Xiao\inst{1}} 

\authorrunning{H. Shi et al.}
%
\institute{Ping An Technology (Shenzhen) Co., Ltd., Shenzhen, China\and
University of Science and Technology of China, Hefei, China}

\maketitle              
\begin{abstract}
Although current Text-To-Speech (TTS) models are able to generate high-quality speech samples, there are still challenges in developing emotion intensity controllable TTS. Most existing TTS models achieve emotion intensity control by extracting intensity information from reference speeches. Unfortunately, limited by the lack of modeling for intra-class emotion intensity and the model's information decoupling capability, the generated speech cannot achieve fine-grained emotion intensity control and suffers from information leakage issues. In this paper, we propose an emotion transfer TTS model, which defines a remapping-based sorting method to model intra-class relative intensity information, combined with Mutual Information (MI) to decouple speaker and emotion information, and synthesizes expressive speeches with perceptible intensity differences. Experiments show that our model achieves fine-grained emotion control while preserving speaker information.

\end{abstract}

\section{Introduction}  
Due to the lack of emotional information, despite the advancements in sequence-to-sequence speech synthesis models \cite{23a}, TTS models often produce high-quality speech that lacks satisfactory expressiveness. It is imperative to resolve this issue, considering the broad range of applications for emotional speech synthesis, particularly in fields such as education and voice assistants. A straightforward method is to use the reference audio to assist emotional speech synthesis. Emotion transfer TTS\cite{CET,23qi} aims to convert speech samples from one speaker to another while preserving the emotional style of the original speech. Moreover, emotional intensity information has also been employed to aid models in achieving emotional control\cite{emomix}.

However, the expressiveness of the generated speech in emotion style transfer heavily relies on the quality of the input reference speech, when aiming to modify the emotion intensity of the output speech while maintaining the speaking style of the source speaker, the only viable approach is to rely on the source speaker to independently control their emotion intensity. For instance, when an individual states, ``I don't want to play with you anymore!'', the sentiment expressed is one of anger. The phrase ``Could you be a little angrier?'' is commonly used to request an increase in the intensity of the emotional expression.
Obviously, it is difficult for ordinary people to grasp the vague vocabulary that describes the intensity such as ``a little bit'' which is not accurately quantified. To promote the effective application of speech synthesis technology among a broader audience, we hope to address the issue of emotion intensity control in the process of emotion transfer. 

In previous intensity controllable Text-to-Speech (TTS) models\cite{mixemo,24md}, researchers have made efforts to extract emotion intensity information from datasets lacking intensity labels. The widely adopted approach is using the relative attributes method\cite{rar} for emotion ranking, which compares non-neutral and neutral speeches to extract emotion intensity information\cite{aslp-fec,aslp-asru}. However, this method follows a premise and assumption: in the same emotional set, the intensity of all emotional speech samples is considered to be the same. This means that the model only measures the distance between emotional speech samples and neutrality, while ignoring the intensity differences among samples within the emotional category. It limits fine-grained modeling of the intra-class emotion intensity, posing a challenge to enhancing flexibility and accuracy in emotion transfer models. 

To address these challenges, we propose a \textbf{R}emapping-based \textbf{S}orting method and construct an \textbf{E}motional \textbf{T}ransfer TTS model (named \textbf{RSET}) to fine-grained extract and control intra-class emotional intensity. Specifically, we design a remapping method to perceive the relative emotional intensity of samples within the same emotional category and perform fine-grained modeling. To perceive the emotional intensity information in the reference audio, an emotional intensity controller is built to perceive and regulate the emotional intensity in the reference speech, and finally obtain the regulated emotional embedding. Furthermore, our approach designs distinct encoders for speaker and emotion information from the reference audio. To refine the model's ability to separate these, we also integrate a loss function that minimizes mutual information \cite{mim}. The resulting distinct speaker and emotion embeddings are subsequently synthesized to guide the TTS model towards producing speech with desired emotion. To ensure the consistency of the speaker information, we formulate a speaker consistency loss by assessing and comparing the speaker information within both the reference and the generated mel-spectrograms. The contributions of our work are summarized as follows:

\vspace{-3pt}
\begin{itemize}
\item We propose an emotion transfer TTS model named RSET which can perform intensity modeling in emotional intra-class space, achieving fine-grained emotion intensity perception and control.
\item We introduce mutual information minimization and speaker consistency loss to improve the decoupling ability of the model while ensuring the invariance of speaker information.
\item Our experiments indicate that the RSET model surpasses two state-of-the-art controllable emotional TTS models.
\end{itemize}
\vspace{-3pt}

\section{Related Works} 
The majority of current emotional speech synthesis systems utilize sequence-to-sequence methods, which were initially explored within the field of machine translation\cite{transformer}, sequence-to-sequence models were subsequently applied to speech synthesis. The structural characteristics of sequence-to-sequence models enable efficient modeling of various time scales (from words to phrases to sentences), showcasing robust long-term dependency modeling capabilities. Through the alignment capability of attention mechanisms, sequence-to-sequence models can capture rhythmic variations in speech details. Additionally, they can predict the duration of speech segments, enabling fine-grained control of prosody. Models based on the sequence-to-sequence architecture usually employ two methods to simulate speech emotion: using explicit labels\cite{21,22} and depending on reference audio\cite{gst,CET}.

Using explicit emotion labels to represent emotion is the most direct method\cite{21,22}, wherein the model is trained to associate particular labels with their corresponding emotional styles. Tits et al.\cite{22} achieved low-resource emotional text-to-speech using models with a small number of emotion labels.

Since emotional speech often lacks a large number of emotion style labels for learning, the most widely used approach in emotional speech synthesis involves transferring emotional styles using reference audio\cite{gst}. Wang et al. \cite{gst} proposed the Global Style Token (GST) as an extension of the reference encoder method, enabling unsupervised learning of style embeddings. By leveraging GST, the model gains the capability to acquire the style from reference audio, facilitating control over synthesized speech style by selecting particular tokens. However, GST lacks generalization ability beyond the training range, which might lead to suboptimal performance when transferring to reference audio outside the training data. Li et al.\cite{CET} introduced a method for emotional speech synthesis within the Tacotron \cite{tacotron} framework. They utilized a reference encoder to infuse emotional embeddings into the synthesis process. Additionally, it uses a style loss function to reduce the stylistic discrepancies between the synthesized and reference mel-spectrograms. Some other studies have also explored the use of Variational Auto Encoders (VAE) to control speech style \cite{vaetts}.

Furthermore, alongside emotional style transfer, controlling emotional intensity has received widespread attention in recent years\cite{emodiff}. Emotional intensity is a complex and subjective aspect of speech that is influenced by various sound cues associated with speech emotions \cite{24a}, making it difficult to analyze and control. To address this challenge, some research studies have utilized additional features such as Voiced, unvoiced, and silence (VUS) states\cite{31}, distance-based quantization\cite{emoqtts} to control emotional intensity. Some works\cite{aslp-asru,aslp-fec} introduce relative attributes\cite{rar} to learn representations of emotional intensity. They automatically learn relationships between low-level acoustic features and different intensity levels of high-level emotional expressions using a relative attribute ranking method, using the obtained ranking results to represent the emotional intensity of emotional samples and guide relevant TTS models (such as FastSpeech2\cite{fastspeech2}) to synthesize emotions of different intensities.

However, the above methods uniformly model different emotional styles, lacking differentiation among samples within the same emotion category. In this paper, based on relative attribute ranking, we utilize a remapping method to model intra-class differences in emotional styles within reference audio, aiming to enhance fine-grained emotional control.

\begin{figure*}[ht]
\begin{minipage}[b]{1.0\linewidth}
  \centering
  \centerline{\includegraphics[width=12cm]{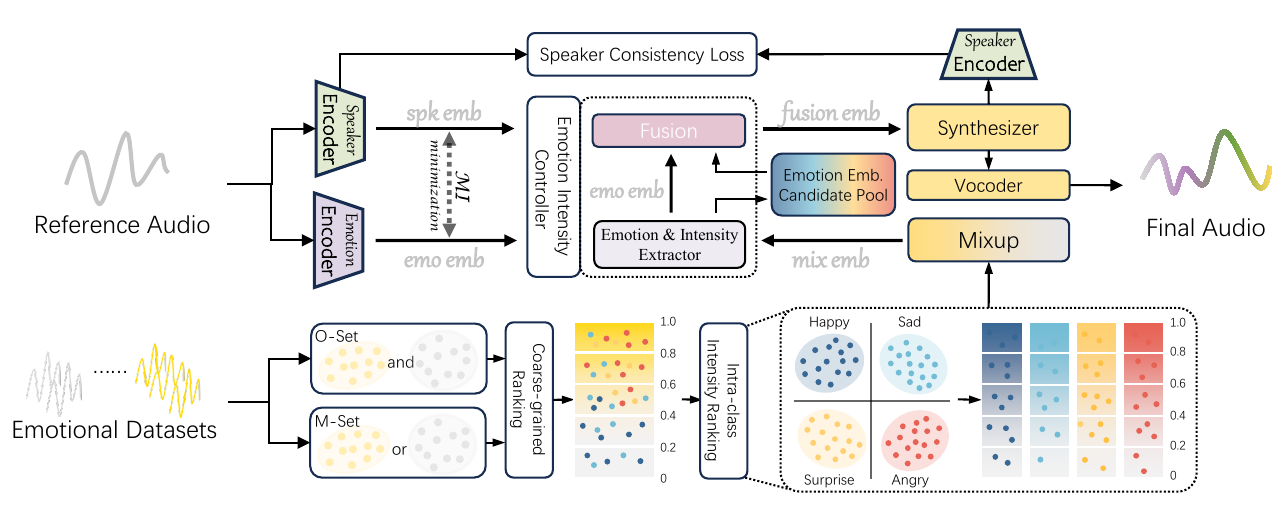}}
\end{minipage}
\caption{The architecture of RSET model. The upper part consists of three modules from left to right: the information decoupling module, the emotion intensity control module, and the synthesizer module. The emotion intensity control module includes an emotion intensity controller and a fusion module. The lower section illustrates the remapping sorting method's overall process, the final sample points represent speech features containing fine-grained intensity information, constituting the emotion-embedding candidate pool for emotion speech synthesis.}
\label{arc}
\end{figure*}

\vspace{-20PT}

\section{Methods}  
\subsection{Overview}
The architecture of the proposed model, as shown in Fig. \ref{arc}, is divided into three parts: the remapping phase based on a ranking model, emotion intensity controller, and information decoupling module. Specifically, the remapping phase is used to extract intra-class intensity labels. The intensity controller is responsible for the control of emotion intensity and outputs the fusion embeddings which guide the TTS model\cite{fastspeech2} to generate the final speech.

\subsection{Remapping-based Sorting Method}
We aim to fine-grained perceive and adjust intra-class emotion intensity. However, attempting to directly map the emotional speech sample space to the intra-class emotion intensity space is not dependable, which inevitably results in mismatches of intra-class intensity. The baseline intensity varies across category spaces, for instance, the intensity of surprise tends to be higher than that of sadness. Hence, it is necessary to roughly model speech intensity in advance. To unsupervisedly extract speech intensity information, we first model a ranking function $r(x_{t})$ in the sample space $\mathbb{R}$ composed of neutral and non-neutral samples:

\begin{equation}
\begin{aligned}
r(x_{t})=\omega x_{t}
\end{aligned}
\end{equation}
where $x_t$ is the audio feature, and $\omega$ is the ranking weight we need to learn. At the same time, following the relative attribute\cite{rar}, we split $\mathbb{R}$ into $O$ and $M$, which are composed of sample pairs with different emotion categories (neutral and non-neutral emotions) and same category (neutral or non-neutral emotion). In order to distinguish and rank different emotions, the ranking function must meet the following criteria: in the O set, emotional intensities of non-neutral emotional samples consistently exceed those of neutral emotional samples, while in the M set, emotional intensities are uniform across all samples. Therefore, the ranking function needs to satisfy the following constraints: 

\begin{equation}
\begin{aligned}
\forall (i,j)\in O:\omega x_i > \omega x_j\\
\forall (i,j)\in M:\omega x_i = \omega x_j
\end{aligned}
\label{OM}
\end{equation}
where $x_i$ and $x_j$ represent two different audio features respectively. While this is an NP hard problem, we adopted the method proposed in \cite{22rar}, which obtained approximate solutions by introducing slack variables $\xi_{i,j}$ and $\gamma_{i,j}$. Subsequently, it becomes necessary to address the subsequent optimization challenge, taking into account the integrated constraint conditions:

\begin{equation}
\begin{aligned}
\underset{\omega}{min} \qquad &(\frac{1}{2} \parallel w \parallel_2^2 + C(\sum \xi _{i,j}^2+\sum \gamma _{i,j}^2)) \\ 
 s.t. \qquad  &wx_i\ge wx_j+1-\xi_{ij};\forall(i,j)\in O\\&|wx_i- wx_j|\le\gamma _{ij};\forall(i,j)\in M\\&\xi_{i,j}\ge0;\gamma_{i,j}\ge0
\end{aligned}
\label{lim}
\end{equation}
where $C$ is employed to regulate the balance between the margin and the size of the slack variables $\xi_{i,j}^2$ and $\gamma_{i,j}^2$.

Assuming that there are a total of $K$ emotional categories (excluding neutral emotions), with each category containing $N$ emotional speech samples. After training the ranking weight $\omega$, we obtain the emotional intensity label $I(E_k,S_n)$ by:
\begin{equation}
\begin{aligned}
I(E_k,S_n)=\omega x_n,\{0\le k \le K;0\le n \le N\}
\end{aligned}
\end{equation}
where $(E_k,S_n)$ represents the $n^{th}$ speech sample feature in the $k^{th}$ emotional category. 

However, the intensity label $I(E_k,S_n)$ is coarse-grained and it only represents the intensity distance of the emotional sample to neutral speech. In order to perceive the intra-class intensity information, we design a quantification function to map the inter-class emotional space to the intra-class emotional space. Specifically, we first calculate the intra-class emotional intensity mean for each emotional category based on the emotion classes:
\begin{equation}
\begin{aligned}
I_{mean}(E_k,\sim)=\frac{1}{N}\sum_{n=1}^N I(E_k,S_n)
\end{aligned}
\end{equation}
Then, we quantify and activate the distance of each sample to the mean intra-class emotional intensity, obtaining the final relative emotional intensity. This process achieves the remapping of the intra-class emotional space:
\begin{equation}
\begin{aligned}
\mathcal{I}_{remap}(E_k,S_n)=sigmoid \left [ I(E_k,S_n)- I_{mean}(E_k,\sim)\right ]  
\end{aligned}
\end{equation}
$\mathcal{I}_{remap}(E_k,S_n)$ represents the final emotional intensity of the sample, which signifies the remapped intra-class emotional intensity.

The detailed procedure for the remapping method is outlined in Algorithm~\ref{algo}.
\begin{algorithm}
\label{algo}
\caption{Remapping-based Sorting Method}
    \KwIn{Emotional speech features inputs $x_t\in \mathbb{R}$ } 
    \KwOut {Intra-class relative intensity value $I_{remap}$}
    
    Initialize $\omega$ with random values, slack variables $\xi_{i,j}$ and $\gamma_{i,j}$ to 0.

    split $\mathbb{R}$ into O set and M set according to Eq. \ref{OM}
    
    train ranking function using Eq. \ref{lim}

    get $n^{th}$ sample intensity value $I(E_k,S_n)$ in emotion class $E_k$ using trained ranking function $\omega$: \\

    $I(E_k,S_n)=\omega x_n,\{0\le k \le K;0\le n \le N\}$
    
    \textbf{for} $k^{th}$ emotion class in samples do:

        \quad total = SUM$(I(E_k,\sim))$

        \quad $N_k$ = length$(E_k)$
        
        \quad ave = total / $N_k$

        \quad \textbf{for} $n^{th}$ sample in emotion class $E_k$ do:
        
        \qquad $I_{remap}$ = $sigmoid(I(E_k,S_n) - ave)$

        \quad \textbf{end for}

    \textbf{end for}
\end{algorithm}

\vspace{-15pt}

\subsection{Emotion Intensity Controller}   
Once the relative intensity labels are obtained, we can learn the intra-class emotion intensity information and perform fine-grained regulation. To this end, we propose an emotion intensity controller, which is composed of emotion intensity extraction and fusion module, as shown in Fig.~\ref{arc}.

Given the limited availability of current emotional datasets, we mix neutral emotion and non-neutral emotions, denote as $E_i$ and $E_j$: 
\begin{equation}
\begin{aligned}
\tilde{E} = \lambda E_i+(1-\lambda )E_j
\end{aligned}
\end{equation}
where $\lambda$ is drawn from a beta distribution randomly, and $\tilde{E}$ denotes the fused emotion embedding. These are then fed into the intensity extraction model. 

Inspired by the excellent performance of NLinear\cite{aaai23} on time series forecasting tasks, we design emotion classfier and intensity extractor to predict emotion class and perceive the intra-class emotion intensity information. The emotion intensity extractor is constructed with an NLinear layer, succeeded by a pair of dense layers and an average pooling layer. The NLinear layer transforms emotion information into latent states, dense layer and global average pooling layer regress the frame-wise hidden states and finally output the utterance-level intensity score $y_{pred}$. Similar to intensity extractor, the emotion predictor consists of an NLinear layer and a softmax layer. The NLinear layer produces emotional features and generates the final emotion prediction result via the connected softmax layer.

\begin{figure}[ht]
\begin{minipage}[b]{1.0\linewidth}
  \centering
  \centerline{\includegraphics[width=12cm]{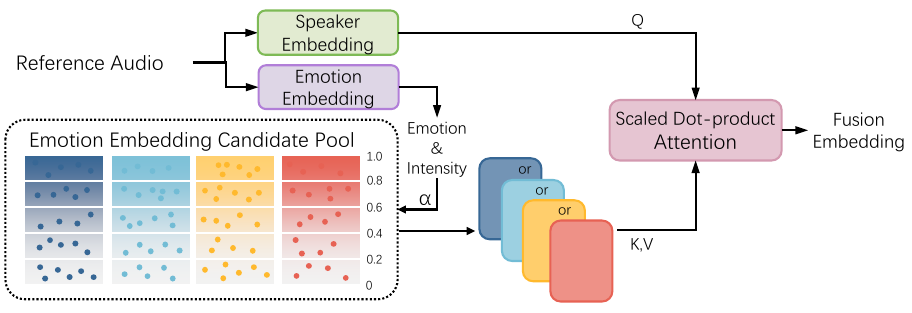}}
\end{minipage}
\caption{Candidate pool and attention fusion module during inference.}
\label{cadi}
\end{figure}

The fusion module utilizes the mechanism of scaled dot-product attention \cite{transformer}, to derive the fusion embedding, and guides the synthesizer in generating emotional speech. We aim to control the emotional intensity by manually setting the intensity adjustment value $\alpha$, but using a scalar alone cannot manage emotional information. Therefore we introduce an emotion embedding candidate pool mechanism, providing candidate emotion embeddings as attention keys and values, as shown in Fig. \ref{cadi}. Utilizing the designated emotional intensity level, we proceed to extract the pertinent emotion-specific embedding. In particular, the collection of emotion embeddings is formed by encoding $N$ distinct representations corresponding to various emotion categories, such as happiness or anger, which are obtained through the remapping-based sorting method, along with their corresponding intra-class emotion intensities. In Fig. \ref{cadi}, the speaker encoder extracts speaker information from the reference audio. Subsequently, the attention block integrates speaker information with emotion embedding information to generate the final fusion embedding.
\vspace{-20pt}
\vspace{-15pt}

\subsection{Information Decoupling and Speaker Consistency}
To mitigate the impact of speaker timbre and emotion information leakage on model performance, we integrate the concepts of mutual information along with the variational contrastive log-ratio upper bound (vCLUB) \cite{mim}, which serves to enhance the model's decoupling capability. Mutual information measures the relationship between speaker information ($I_s$) and emotion information ($I_e$), vCLUB restricts the MI between $I_s$ and $I_e$, thereby aiding the model in achieving comprehensive decoupling of the two:
\begin{equation}
\begin{aligned}
I_{vCLUB}(I_s,I_e)=&\mathbb{E}_{p(I_s,I_e)}[\log {q_{\theta }(I_e|I_s)}]\\
&-\mathbb{E}_{p(I_s)}\mathbb{E}_{p(I_e)}[\log {q_{\theta }(I_e|I_s)}]
\end{aligned}
\end{equation}
where variational distribution $q_{\theta }(I_e|I_s)$ with parameter $\theta$ aims to approximate the actual probability $p(I_e|I_s)$. Similar to \cite{vqmivc}, we use unbiased estimation for vCLUB between $I_s$ and $I_e$ as the loss function of MI minimization:

\begin{equation}
\begin{aligned}
\mathcal{L}_{MI} = \frac{2}{N^2} \sum_{n=1}^{N} \sum_{k=1}^{N} \log \left [ {q_{\theta }(I_{e,n}|I_{s,n})}-{q_{\theta }(I_{e,k}|I_{s,n})} \right]
\end{aligned}
\end{equation}
where N is the number of utterances, $I_{e,i}$ and $I_{s,i}$ represent the speaker and emotion information of the $i^{th}$ sample respectively.

Meanwhile, in order to ensure the consistency of speaker identity, we introduce a novel loss function, denoted as $\mathcal{L}_{Spcon}$ designed to reduce the disparity in the L2 norm between the speaker information present in the synthesized and the reference speeches:
\begin{equation}
\begin{aligned}
\mathcal{L}_{Spcon} = \frac{1}{N} \sum_{n=1}^{N}  \left ( \hat{I}_{s,i} - I_{s,i}\right)^2
\end{aligned}
\end{equation}
where $\hat{I}_{s,i}$ represents the speaker information of the output audio feature, and $I_{s,i}$ represents the speaker information of the reference audio.

\subsection{Loss Function}
The loss function within our proposed model encompasses three distinct components: reconstruction loss $\mathcal{L}_{Recon}$, MI minimization loss $\mathcal{L}_{MI}$ and speaker consistency loss $\mathcal{L}_{Spcon}$. Reconstruction loss is is quantified by the L1 norm discrepancy between the mel-spectrogram of the target audio and that of the synthesized. Finally, the total loss of RSET is:  
\begin{equation}
\begin{aligned}
\mathcal{L}_{RSET} = \mathcal{L}_{Recon} + \alpha_1 \mathcal{L}_{MI} + \alpha_2 \mathcal{L}_{Spcon}
\label{loss}
\end{aligned}
\end{equation}
where $\alpha_1$ and $\alpha_2$ are the weights of the corresponding losses respectively.

\subsection{Training and Inference}
\textbf{Training}: our model's training is divided into a two-phase process. The initial phase is dedicated to training the remapping-based ranking module, which is used to construct the emotion embedding candidate pool. In the second phase, we train the overall speech synthesis network. During the training of the second stage, since emotion control relies on emotion information encoding, we first jointly train two encoders and the synthesizer. Subsequently, we deploy the already trained emotion encoder to distill emotional cues from the reference audio, which is then utilized to further train the emotion intensity extractor. Finally, the modules from the second stage are combined and used during the inference stage.\\
\textbf{Inference}: in the inference stage, we follow the description of the emotion intensity control section, as illustrated in Fig. \ref{cadi}. By matching the adjusted intensity value in the candidate pool, we guide the synthesizer to generate audio corresponding to the desired emotional intensity. Specifically, we first use the emotion intensity extractor to predict the emotion category and intensity information of the reference audio. Then, we combine the intensity value with the control value $\alpha$ to generate the final adjusted intensity value. The emotion category and intensity value help index emotion embedding, which are then fused with speaker information to guide the generator in producing the ultimate emotional speech.

\section{Experiments}

\subsection{Experimental Setup}
All experiments were performed on the English subset of the Emotional Speech Dataset (ESD) \cite{ESD}. This subset consists of recordings from 10 speakers, each expressing five emotional categories: anger, happiness, sadness, surprise, and neutrality. Each speaker provided 350 parallel utterances for each emotion category, resulting in approximately 1.2 hours of speech per speaker.

To evaluate our model, we selected the following methods to compare with our RSET:
\begin{itemize}
\item \textbf{GT} and \textbf{GT(voc.)}: ground-truth audios and samples generated by HiFi-GAN \cite{hifi} based on extracted ground truth mel-spectrogram.
\item \textbf{Mixed Emotion}: proposed in\cite{mixemo}, it uses relative attribute sorting to learn emotion intensity, thus achieving emotion intensity control.
\item \textbf{EmoMix}\cite{emomix}: based on the diffusion model, it can generate speeches with a specified emotional intensity by mixing neutral and target emotion in different proportions. 
\end{itemize}
Notably, in quality and similarity evaluation and ablation study, the samples from RSET, Mixed Emotion, and EmoMix were controlled under an intensity weight of $\alpha = 1.0$. This allows them to be directly compared with other samples. Additionally, to illustrate the significance of our subjective experiments, we computed the mean and variance of all results. Considering the corresponding sample sizes, we performed t-test on the experimental results. The results consistently showed that, at a significance level of $\alpha$ = 0.05, the p-value with Mixed Emotion were below 0.05, except 0.1 for EmoMix, meeting a confidence level of 95$\%$ and 90$\%$.

\subsection{Implementation Details}
In the remapping-based sorting method, we initially use the Opensmile toolkit\cite{opensmile} to extract acoustic features from speech samples, including energy, F0, and others. These extracted acoustic features are then used to derive emotional intensity information. 
Additionally, we leverage mutual information to enhance the information decoupling capabilities of the two encoders, following the implementation in \cite{vqmivc}. The synthesizer adopts a structure similar to Fastspeech2. The speaker encoder and emotion encoder adhere to the structure proposed by mel-style encoder\cite{metaspeech}. In the emotional intensity controller, we feed the 256-dimensional emotional embedding into NLinear. According to \cite{aaai23}, the look-back window size of NLinear for both emotion and intensity extractor is set to 96. As for training, we utilized a batch size of 32 and the ADAM optimizer applied with 1e-4 learning rate. In Eq.\ref{loss}, throughout all our experiments, we consistently assigned the values of $\alpha_1$ and $\alpha_2$ to be 0.1 refering to \cite{vqmivc}. The two stages were trained for 1M and 25k steps, respectively.

\begin{table}[h]
\centering
\caption{Subjective and objective experimental results on ESD dataset.}
\renewcommand{\arraystretch}{1.1}
        \begin{tabular}{p{3cm} | p{2.2cm}<{\centering}  p{2.2cm}<{\centering}  p{2.2cm}<{\centering} | p{1.3cm}<{\centering}}
        \hline
         Model & \textbf{MOS}$\uparrow$ & \textbf{SMOS}$\uparrow$ & \textbf{EmoAcc}$\uparrow$ & \textbf{MCD}$\downarrow$  \\
        \hline
        GT & 4.49 $\pm$ 0.08 & / & / & / \\
        GT (voc.) & 4.41 $\pm$ 0.07 & 4.48 $\pm$ 0.06 & 99.63\% & 2.40 \\
        \hline
        Mixed Emotion\cite{mixemo} & 3.73 $\pm$ 0.12 & 3.85 $\pm$ 0.09 & 98.79\% & 5.03 \\
        EmoMix\cite{emomix} & 3.89 $\pm$ 0.08 & 3.92 $\pm$ 0.08 & 98.45\% & \textbf{4.79} \\
        \hline
        Ours & \textbf{3.98 $\pm$ 0.06} & \textbf{4.15 $\pm$ 0.07} & \textbf{99.31\%} & 4.81 \\
        \hline
        \end{tabular}
\label{mos}
\end{table}

\vspace{-20pt}
\subsection{Quality and Similarity Evaluation}
The synthesis quality, speaker consistency, and emotional precision of the generated speech were evaluated through a combination of subjective and objective experimental methods. For subjective assessment, we invited 30 native speakers to participate in both Mean Opinion Score (MOS) and Similarity Mean Opinion Score (SMOS) evaluations. During each evaluation, the participants were tasked with scoring the set of five speeches per emotion category using a numerical scale ranging from 1 to 5, with increments of 1 point. Furthermore, they were required to carry out an emotional classification task on the speech samples to assess the emotional accuracy. The outcomes of these assessments are presented in Table \ref{mos}. RSET achieves the highest MOS, surpassing the Mixed Emotion by 0.25. Thanks to information decoupling and speaker invariance constraints, our model also achieves the highest SMOS score. Besides, the samples generated by our model exhibit the highest emotional recognition accuracy, which means that RSET can generate accurate emotional speech. For the objective evaluation metric, we utilize Mel-Cepstral Distortion (MCD)\cite{mcd} to measure the similarity between the original and synthesized Mel-spectrograms. The results shows our RSET scores only 0.02 points lower than EmoMix, but outperforms Mixed Emotion by 0.22. The experimental results indicate that RSET enhances both speaker and emotion similarity while maintaining speech quality. Audio samples are available at \url{https://lemon-ustc.github.io/RSET-demo}.

\begin{figure*}[t]
\centering  
\subfigure[Happy]{
\includegraphics[width=0.45\columnwidth]{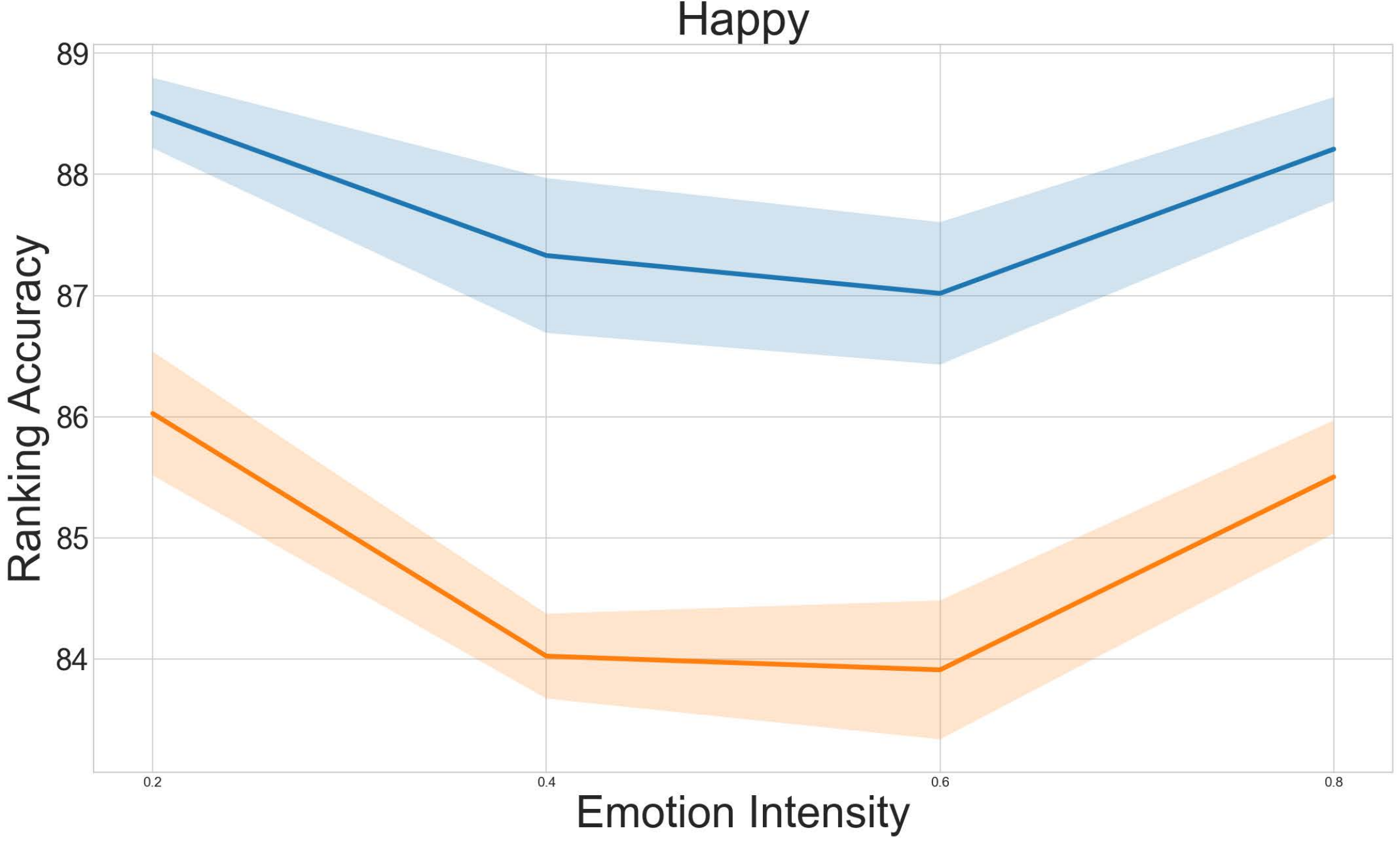}} 
\subfigure[Sad]{
\includegraphics[width=0.45\columnwidth]{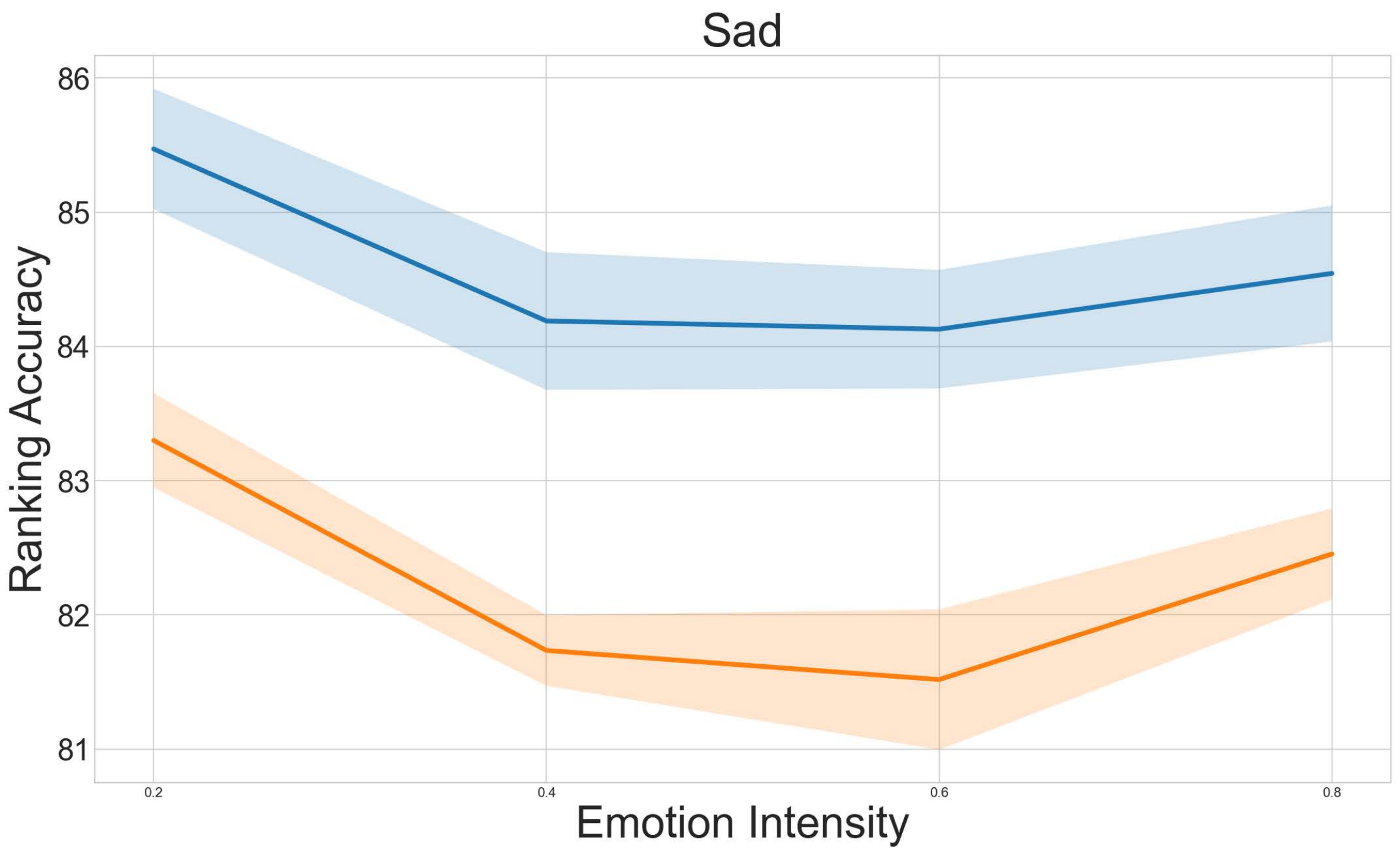}} 
\subfigure[Surprise]{
\includegraphics[width=0.45\columnwidth]{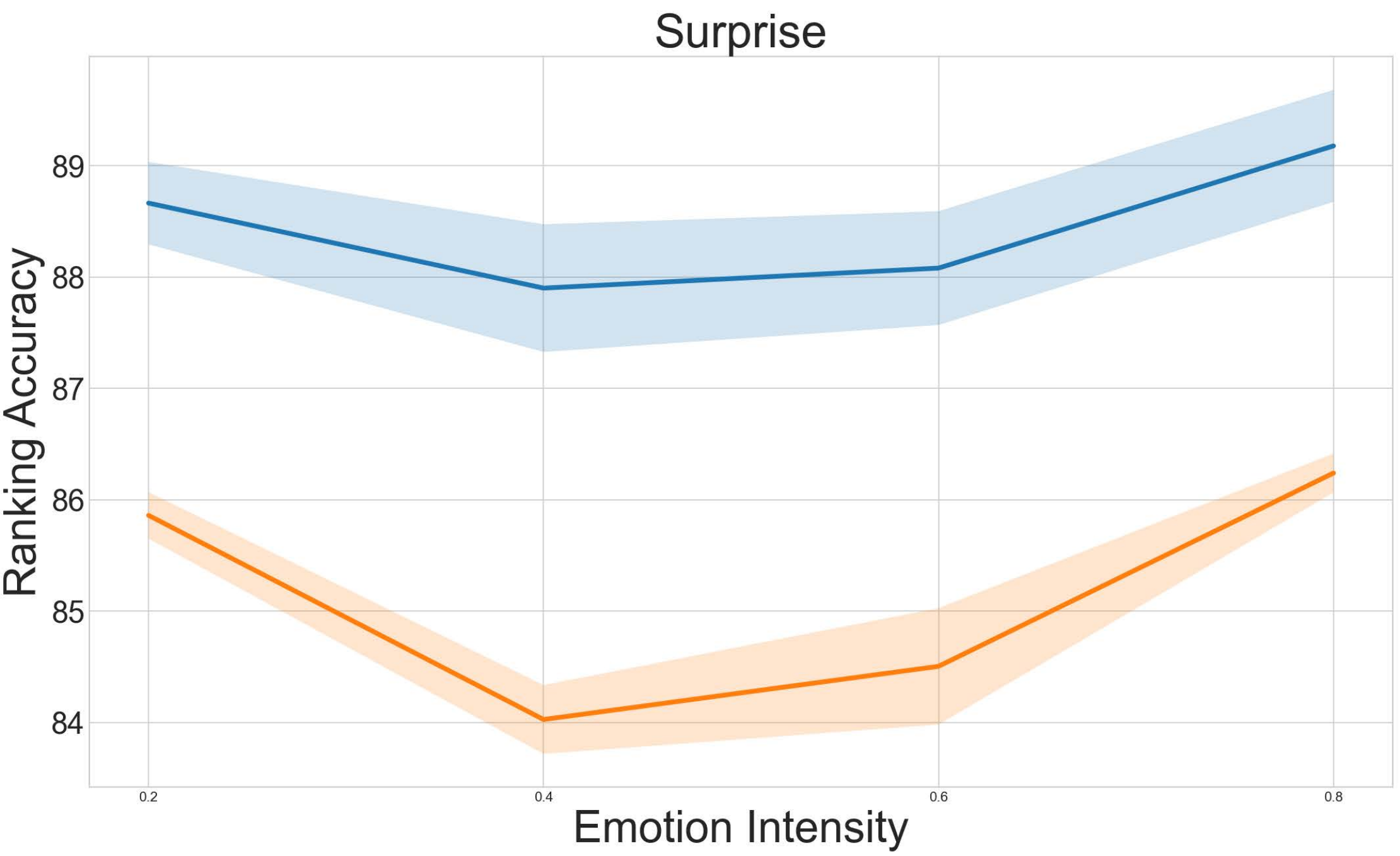}}
\subfigure[Angry]{
\includegraphics[width=0.45\columnwidth]{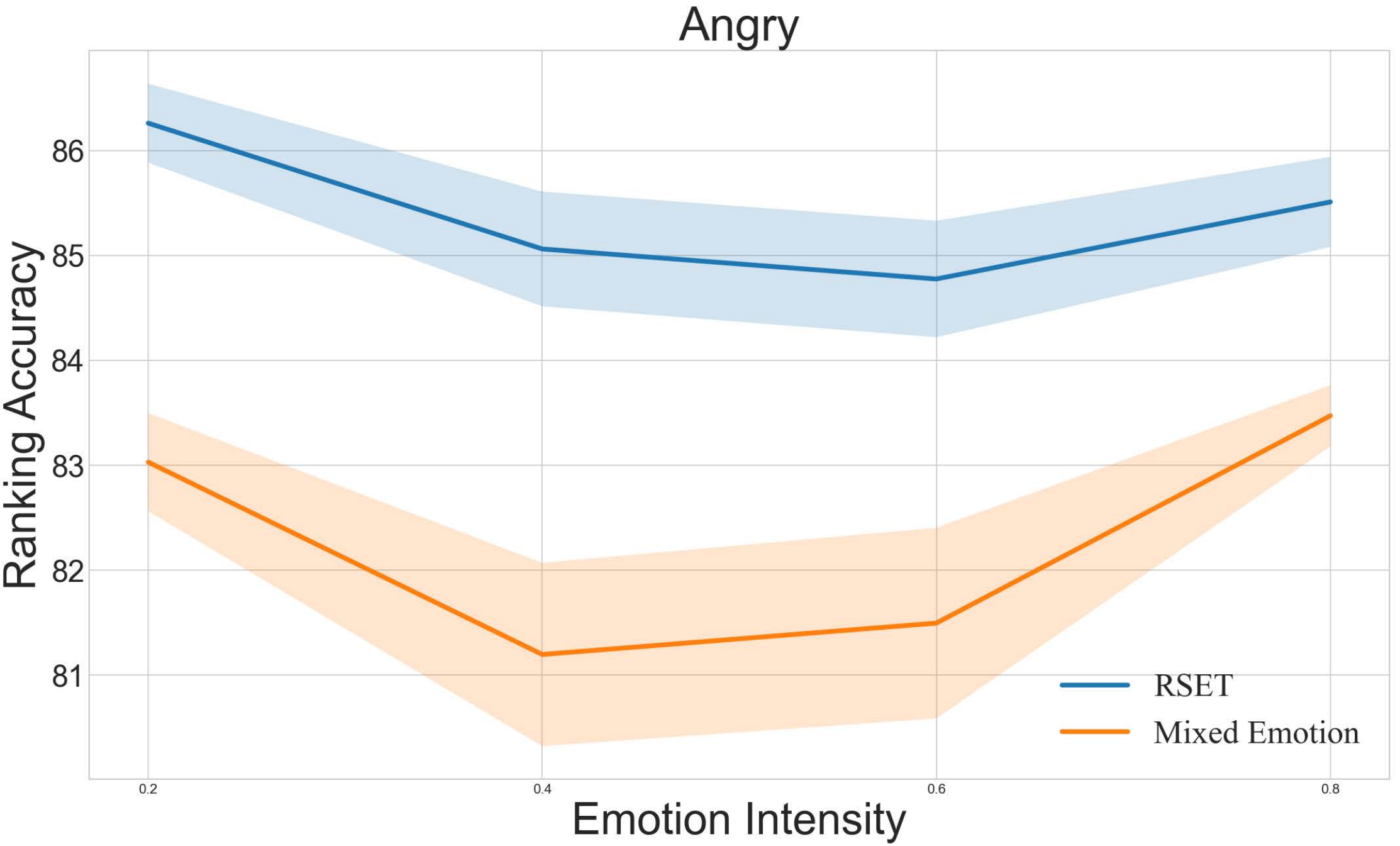}}
\caption{Comparison curve of emotion sorting accuracy. The straight line in the center indicates the mean accuracy for each intensity level, with lighter hues on both sides representing the variance. Blue represents RSET, while orange corresponds to Mixed Emotion.}
\label{control}
\end{figure*}

\begin{table}[h]
    \centering
    \caption{MOS comparison for ablation study.}
    \renewcommand{\arraystretch}{1.1}
        \begin{tabular}{p{4cm}<{\centering} | p{2cm}<{\centering} | p{2cm}<{\centering}}
        \hline
         & \textbf{MOS}$\uparrow$ & \textbf{SMOS}$\uparrow$ \\
        \hline
        Ours & / & / \\
        w/o Remap & -0.18 & -0.05 \\
        w/o MI minimization & -0.08 & -0.06 \\
        w/o Consistency loss & -0.05 & -0.09 \\
        \hline
        \end{tabular}
\label{ab}
\end{table}
\vspace{-10pt}

\subsection{Ablation Study}
Ablation studies were performed to individually remove the remapping module, MI, and the consistency loss separately to demonstrate the effectiveness of our proposed modules. In this section's experiments, we still used the MOS and SMOS metrics to measure the expressiveness of the synthesized speeches, as indicated in Table \ref{ab}. To validate the effectiveness of the remapping method, we first removed the remapping module, extracting intensity information solely using the basic ranking method. The results showed a decrease of 0.18 in MOS, preliminary evidence that our remapping method improves the overall expressiveness of emotional speech, while SMOS remained almost unchanged, as the remapping method does not involve speaker information and therefore has no impact on result similarity. Next, we removed the mutual information constraint between the two decoders in the reference audio decoupling part. This led to a decrease of 0.08 in MOS and 0.06 in SMOS, indicating that the MI minimization module disentangled information in the reference speech during the information transfer process, separating speaker information from emotional information. Additionally, we attempted to exclude the speaker consistency loss during the training process. The significant decrease in speaker similarity scores indicates that the consistency loss constrains the offset of speaker information, allowing the model to accurately output the target speaker's speech.

\begin{figure}[t]
\centering  
\subfigure[Happy]{
\includegraphics[width=0.36\columnwidth]{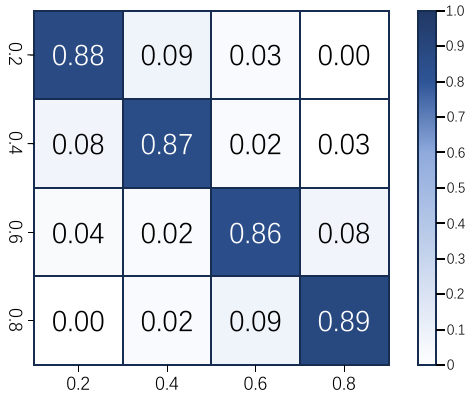}} 
\subfigure[Sad]{
\includegraphics[width=0.36\columnwidth]{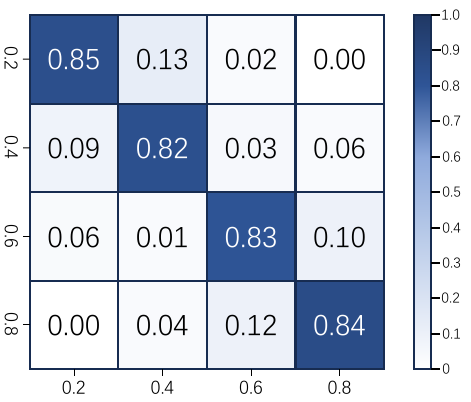}} 
\subfigure[Surprise]{
\includegraphics[width=0.36\columnwidth]{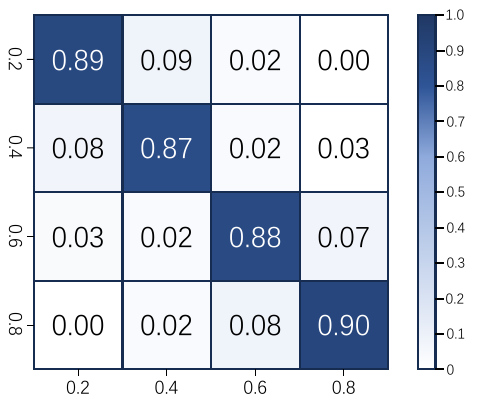}}
\subfigure[Angry]{
\includegraphics[width=0.36\columnwidth]{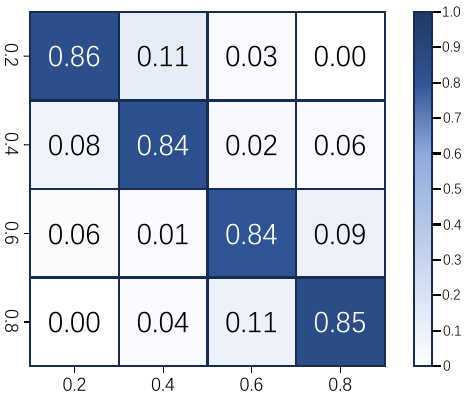}}
\caption{The confusion matrix for different emotion intensity results. It utilizes the horizontal axis to depict the actual stages of intensity values, while the vertical axis corresponds to the artificially sorted results.}
\label{matrix}
\end{figure}
\vspace{-10pt}

\subsection{Emotion Intensity Controllability}
\begin{figure}[ht]
\begin{minipage}[b]{1.0\linewidth}
  \centering
  \centerline{\includegraphics[width=0.8\columnwidth]{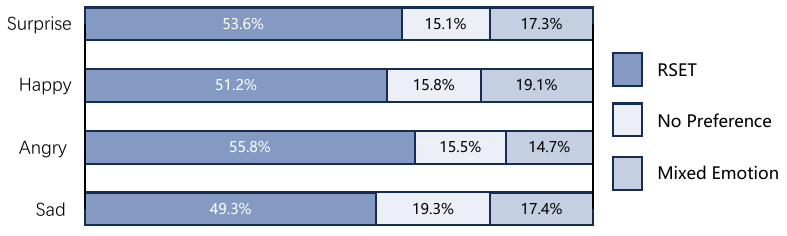}}
  \caption{A/B preference test for RSET and Mixed Emotion.}
  \label{tab:prefer}
\end{minipage}
\end{figure}

In order to evaluate the emotion intensity control ability of our proposed model, we conducted a subjective assessment on generated speech samples with different emotional intensities. For ease of participant scoring, we selected four intensity value stages (0.2, 0.4, 0.6, 0.8) from (0,1). Participants were required to compare and rank the synthesized speech samples concerning the reference audio based on their perceived intensity differences. We collected the ranking results from each participant, calculated the accuracy for each stage, and created a comparative curve with Mixed Emotion as shown in Fig. \ref{control}. RSET achieved leading accuracy in almost all stages, indicating that our proposed model generates emotion speech with intensity that is more easily perceptible. Then we generated a confusion matrix according to the ranking results, as depicted in Fig. \ref{matrix}. In the confusion matrix, the vertical and horizontal coordinates of each sub-graph represent the actual sorting of emotional intensity and the perceived sorting by evaluators, respectively. Ideally, if the participant's ranking matches the actual ranking, the diagonal values of the confusion matrix should all be 1.0. Consequently, larger values along the diagonal of the confusion matrix are desirable, while smaller values elsewhere are preferable. Our model's diagonal scores significantly exceeded the rest, surpassing 0.8 on the diagonals for each emotion. Additionally, values in other cells closer to the edges are smaller, with some extreme positions reaching 0. This demonstrates that the participants could distinctly differentiate between emotional intensity variations in the generated speech by our model. Furthermore, this confirms the effectiveness of our remapping method in discerning differences in emotional intensities within categories, thereby enabling finer control over emotional intensity. Meanwhile, in order to compare the emotional expressiveness of other models, we conducted an A/B preference experiment. Fig. \ref{tab:prefer} shows the preference results, we perform well in emotion expression, outperforming Mixed Emotion.

\vspace{-20pt}

\section{Conclusion}
In this paper, we propose RSET, an emotion transfer model with a remapping method which helps to capture intra-class emotion relative intensity and enable fine-grained control. The intensity controller combines the fine-tuned emotion information with speaker representation decoupling by mutual information minimization and then guides the TTS model to generate high-quality emotional speech. Experimental results show that RSET effectively control emotion intensity while maintaining the high quality of generated speech. 

\section{Acknowledgement}
This paper is supported by the Key Research and Development Program of Guangdong Province under grant No.2021B0101400003. Corresponding author is Xulong Zhang (zhangxulong@ieee.org) from Ping An Technology (Shenzhen) Co., Ltd..

%
%
%
%
\bibliographystyle{splncs04}
\bibliography{z_ref.bib}
\end{document}